\documentstyle[aps]{revtex}
\begin{document}
\draft
\def\be{\begin{equation}}
\def\ee{\end{equation}}
\def\ba{\begin{eqnarray}}
\def\ea{\end{eqnarray}}
\def\bq{\begin{quote}}
\def\eq{\end{quote}}
\def\PL{{ \it Phys. Lett.} }
\def\PRL{{\it Phys. Rev. Lett.} }
\def\NP{{\it Nucl. Phys.} }
\def\PR{{\it Phys. Rev.} }
\def\MPL{{\it Mod. Phys. Lett.} }
\def\IJMP{{\it Int. J. Mod .Phys.} }
\newcommand{\labell}[1]{\label{#1}\qquad_{#1}} 
\newcommand{\labels}[1]{\vskip-2ex$_{#1}$\label{#1}} 
\twocolumn[\hsize\textwidth\columnwidth\hsize\csname
@twocolumnfalse\endcsname \preprint{SU-ITP-99/6\\ hep-th/9904120\\
April 1999}
\date{April 1999}
\title{\Large \bf Cosmology vs. Holography}
\author{Nemanja Kaloper and Andrei Linde}
\address{
Department of Physics, Stanford University, Stanford, CA 94305-4060,
USA } \maketitle
\begin{abstract}
The most radical version of the holographic principle asserts that
all information about physical processes in the world can be stored
on its surface. This formulation is at odds with inflationary
cosmology, which implies that physical processes in our part of the
universe do not depend on the boundary conditions. Also, there are some
indications that the radical version of the holographic theory in  
the context
of cosmology may have problems with unitarity and causality. Another
formulation of the holographic principle, due to Fischler and
Susskind, implies that the entropy of matter inside the
post-inflationary particle horizon must be smaller than the area of
the horizon. Their conjecture was very successful for a wide class of
open and flat universes, but it did not apply to closed universes.
Bak and Rey proposed a different holographic bound on entropy which
was valid for closed universes of a certain type. However, as we will
show, neither proposal applies to open, flat and closed universes
with matter and a small negative cosmological constant. We will
argue, in agreement with Easther, Lowe, and Veneziano, that whenever
the holographic constraint on the entropy inside the horizon is
valid, it follows from the Bekenstein-Hawking bound on the black hole
entropy. These constraints do not allow one to rule out closed
universes and other universes which may experience gravitational
collapse, and do not impose any constraints on inflationary
cosmology.

\end{abstract}
 \vskip2pc]
\section{Introduction}

Recently a new set of ideas was put forward, which was called ``the
holographic principle'' \cite{thooft,lenny1}.
According to this set of ideas,
under certain conditions all
the information about
a physical system is coded on its boundary, implying that the entropy
of a system cannot exceed its boundary area in Planck units.

This principle was motivated by the well-known result
in black
hole theory:  the total entropy $S_m$ of matter inside of a black hole
cannot be greater than the Bekenstein-Hawking entropy, which is
equal to a quarter of the area of the  event  horizon in the Planck
units, $S_m \leq S_{BH} = {A  \over 4}$  \cite{bekenstein}. One can
interpret
this result as a statement that all the information about the interior of
a black hole is stored on its  horizon.

The main aim
of the holographic principle is to extend this
statement to a broader class of situations. This principle, in its
most radical form, would imply that our world
is two-dimensional in a certain sense, because all the
information about physical processes in
the world is stored at its surface.
This conjecture is very interesting, and physical implications of
its most radical version could be quite significant. There has been a lot
of activity related to the use of the holographic principle in quantum
gravity, string theory and M-theory.
For example, there is a
conjecture that the knowledge of a supersymmetric
Yang-Mills theory at the boundary of an Anti-de-Sitter space
may be sufficient
to restore
the information about
supergravity/string theory in the bulk \cite{malda}.

However, if one tries to apply  the holographic principle to
cosmology, one
immediately recognizes several problems. For example, a
closed universe has finite size, but it does not have any boundary.
What is the
meaning of the holographic principle in such a case?
If the universe is infinite (open or flat), then it does not have
boundaries either.
In these cases,
one may try to  compare the entropy inside of a  box
of size $R$ with its area,
and then take the limit as $R \to \infty$. But in this limit  the
entropy is
always larger than the area \cite{lenny3}.

Another possibility is to compare the area of a domain of the size of
the particle
horizon (the causally connected part of the universe) with the entropy
of matter
inside this domain.
But this is also problematic. The entropy produced during reheating after
inflation is proportional to the total volume of inflationary
universe. During
inflation, the  volume inside the particle horizon grows as
$e^{3Ht}$, whereas
the area of the  horizon grows as $e^{2Ht}$.  Clearly, the entropy
becomes much
greater than the area of the horizon if the duration of
inflation is sufficiently large. This means that an inflationary
universe is not two-dimensional; information stored at its
``surface'' is not rich enough to describe physical processes in its
interior. In fact, one of  the main advantages of inflation is the
possibility to study each  domain of size $H^{-1}$ as an independent part
of the universe, due to the no-hair theorem for de Sitter space.
This makes the
events at the boundaries of an inflationary universe irrelevant for
the description of local physics \cite{book}.
Thus, the most radical version of the holographic principle seems to
be at
odds with inflationary
cosmology.

One may try to formulate a weaker form of this
principle, which may still be quite useful. For example,  Fischler and
Susskind proposed to put constraints only on the part of the entropy
which passed through the backward light cone \cite{lenny3}. This
formulation
does not confront inflationary cosmology because it eliminates from the
consideration most of the entropy produced inside the light cone
during the
post-inflationary reheating of the universe. They further concentrated on
investigation of those situations where
cosmological evolution is adiabatic. From the point of view of
inflationary
cosmology, this means that they considered  the evolution of the
universe after
reheating. The largest domain in which all of the entropy crossed
the boundary
when the evolution is adiabatic is bounded by   the light cone
emitted {\it
after} inflation and reheating.  In what follows we will loosely call this
light cone of size $O(H^{-1})$ ``particle horizon,'' even though the true
particle horizon, describing the light cone emitted at the beginning of
inflation,  is exponentially large.

Fischler and
Susskind  argued that in the
case of adiabatic evolution the total entropy of matter within
the particle horizon must
be smaller than the area of the horizon, $S \lesssim A $
\cite{lenny3}.
This conjecture is
rather nontrivial. Indeed, the origin
of the  Bekenstein-Hawking constraint on the entropy of a black hole
is the existence of the  {\it event}  horizon, which serves as a natural
boundary for all processes inside a black hole. But there is no event
horizon in a non-inflationary universe, and the idea to replace it by
the {\it particle} horizon requires some justification. Also, the
Bekenstein-Hawking constraint on the entropy is valid even if the
processes
inside a black hole are non-adiabatic.  Thus it would be desirable to
investigate this proposal and find a way to apply it to the
situations when the
processes can be non-adiabatic.

Remarkably, Fischler and
Susskind  have shown that their conjecture is  valid for a
flat universe
with all possible equations of state satisfying the condition
$0\leq  p  \leq \rho$.
This result suggests that there may be some deep
reasons for the validity of holography. However, they also
noticed that
their version of the holographic principle is violated in a closed
universe.
One may consider this observation
either as an indication that closed universes are
impossible or as a warning,
showing that the holographic
principle may require additional
justification and/or reformulation. Indeed, this
principle is not a rigid scheme but a theory
in the making. It may be quite successful in many respects, but  one
should not be surprised to see some parts of its formulation change.
For example, Bak and   Rey  suggested to replace the particle
horizon by an
apparent horizon in the formulation of the holographic principle,
claiming that their proposal does not suffer from any problems in
the closed universe case \cite{brtwo}.

There were many attempts to apply various formulations of the holographic
principle to various cosmological models, but the existing literature on
cosmic holography is somewhat controversial. The entropy of the
observed
component of matter (such as photons) is well below $10^{90}$
\cite{book}. Meanwhile the constraint $S \lesssim A$
applied to our part of the universe  implies that  $S < 10^{120}$
\cite{lenny3}, which does not look particularly restrictive.
Holography could be quite
important if it were able to rule out
some types of cosmological models,
but this possibility depends on the
formulation and the range of validity of
the holographic principle. One may try
to use holography to solve the cosmological constant problem
\cite{banks,cohen}, but the progress
in this direction was very limited.
Recently it was claimed that holography puts strong constraints on
inflationary theory \cite{infl}, but the authors
of Ref. \cite{el}
argued that
this is not the case. Holographic
considerations were
used in investigation of the
pre-big bang theory \cite{br,bmp,gv},
and on the basis of this investigation
it was claimed that this theory solves
the entropy
problem in the pre-big bang theory\cite{gv}, which is at odds
with the results of  \cite{klb}.

The main goal of this paper is to
examine the basic assumptions of cosmic
holography and check which of them may require
modifications. We will try to
find out whether holography indeed puts
constraints on various cosmological
models. We will show, in particular,
that the original formulation of
the holographic principle should be reconsidered
more generally, and  not only when applied
to closed universes. The holographic entropy bound proposed in
\cite{lenny3}, as well as the
formulation proposed in \cite{brtwo}, is
violated at
late stages of evolution of
open, flat and closed universes
containing usual matter and a small
amount of negative vacuum energy density. At the
beginning of their evolution, such universes
cannot be distinguished from the universe with a positive or
vanishing vacuum energy density. Thus there is no obvious reason to
consider
such universes unphysical and rule them out.
However, when the density of
matter becomes diluted by expansion, a universe with a negative
vacuum energy collapses, and the condition
$S \lesssim A $ becomes violated long before the
universe reaches the Planck density.

 The investigation of universes with a negative cosmological constant
gives  an additional reason to look for a
reformulation of the cosmological
holographic principle.
Our approach will be most closely related to the approach outlined by
Easther and Lowe \cite{el}, and by Veneziano \cite{gv}. They argued
that the entropy of the interior of a domain of
size $H^{-1}$ cannot be greater than
the entropy of a black hole of a
similar radius. We will
extend their discussion and propose a
justification for the entropy bound obtained in Ref.
\cite{lenny3} for the case of an expanding noninflationary (or
post-inflationary) universe.
We will argue, in agreement with \cite{el,gv}, that in those
cases where the
holographic bound of Ref.
\cite{lenny3} is valid, it is equivalent to the Bekenstein-Hawking
bound, which does not require any assumptions about adiabatic evolution.
This bound alone cannot resolve the entropy problem for
the pre-big bang cosmology and does not lead to any constraints on
inflation.

\section{Cosmology and holography}

\subsection{Flat universe with $p =  \gamma \rho$}

Let us begin with a brief review of \cite{lenny3}. We will restrict
our attention to the case when gravitational dynamics is given by
the Einstein's equations, and the evolution is adiabatic. First we
will consider flat homogeneous
and isotropic FRW universes, whose metric is
\begin{equation}
ds^2 = - dt^2 + a^2(t) \left({dr^2} + r^2 d \Omega \right) \ .
\label{metric}
\end{equation}
We will use the units $8\pi{G_N}=1$. For
simplicity we will consider matter with the energy-momentum tensor
$T_{\mu\nu}$ =  diag$(\rho,p,p,p)$. The independent equations of
motion are
\begin{equation}
H^2  = \rho/3 \ ,~~~~~~~~ \dot \rho + 3H (\rho + p) =0 \ ,
\label{eoms}
\end{equation}
where $H = \dot a/a$ is the Hubble parameter, $\rho$
and $p$ are the energy density and pressure, and the overdot denotes
the time derivative. We will assume that $\rho > 0$, $p = \gamma
\rho$, and
that
the energy-momentum tensor satisfies the dominant energy condition
$|\gamma| \leq 1$. This will generalize the results of \cite{lenny3}
obtained
for $0\leq \gamma \leq 1$,
and is in fact the correct sufficient condition
for the validity of the holographic bounds in flat
and open
FRW universes.

The solutions of (\ref{eoms}) for $\gamma > -1$ can be written
as
\begin{equation} a(t) =  t^{\frac{2}{3(\gamma+1)}}\ .
\label{solscale}
\end{equation}
Here we took by definition $a = 1$ at
the Planck time $t = 1$. Density decreases as $ \rho =
\frac{\rho_0}{a^{3(\gamma+1)}}$, where $\rho_0 = {4\over 3 (\gamma +
1)^2}$ is the
density at $t = 1$.
(For $\gamma =-1$ one has the usual de
Sitter solution.) The particle horizon is defined by the distance
covered by the light cone emitted at the singularity $t = 0$:
\begin{equation}
\label{horizon} L_H(t) = a(t) \int^t_{0} \frac{dt'}{a(t')} = a(t)
r_H(t) \ ,
\label{parthor}
\end{equation}
where $r_H$ is the comoving size of the
horizon defined by the condition ${dt\over a } = dr_H$. Suppose
first that
$\gamma > -1/3$. Then the comoving
horizon is
\begin{equation} r_H = L_H/a = {3(\gamma+1)\over 3\gamma +1}~
t^{3\gamma+1\over 3(\gamma+1)} \ ,
\label{comhorizon}\end{equation}
and
\begin{equation}  L_H  = {3(\gamma+1)\over 3\gamma +1} t = {2\over
3\gamma
+1} H^{-1} \ .
\label{comhorizon3}\end{equation}
At the Planck time $t = 1$ one has $ L_H  = {3(\gamma+1)\over3\gamma
+1}$
which generically is $O(1)$. The volume of space within the
distance $L_H$ from any point was also $O(1)$. The entropy density at
that time could not be greater than $O(1)$, so one may say that
initially $\left({S\over A}\right)_0=\sigma \lesssim 1$. Later the
total entropy inside the horizon  grows as
$\sigma L_H^3/a^3$, whereas the
total area $A$ of the particle horizon grows as $L_H^2$.
Therefore
\begin{equation} {S\over A} \sim \sigma {L_H\over a^3} =
\sigma {r_H\over a^2} \ .
\label{comhorizon2}\end{equation}
This yields
\begin{equation}
{S\over A} \sim   \sigma t^{\gamma-1\over
\gamma+1}\ .
\label{comhorizon2a}
\end{equation}
Thus the ratio ${S\over
A}$ does not increase in time for $1\ge \gamma > -1/3$,
so if the holographic
constraint ${S\over A}\lesssim 1$ was satisfied at the Planck time,
later on it will be satisfied even
better \cite{lenny3}.

A similar result can be obtained for  $-1 \leq \gamma \leq -1/3$.
However,
investigation of this case involves several subtle points. First of
all, in
this case the integral in Eq. (\ref{parthor}) diverges at small $t$.
This is
not a real problem though. It is resolved if one defines the
particle horizon
as an integral not from $t = 0$, but from the Planck time $t = 1$.

A more serious issue is the assumption of adiabatic expansion of the
universe. If one makes this assumption, then one can show that the
holographic
bound is satisfied for all $\gamma$ in the interval $-1\le \gamma
\le 1$, which
generalizes the result obtained in \cite{lenny3}.  However, the
universe with
$1+\gamma \ll 2/3$
(i.e. with $\gamma \approx -1$) is inflationary.
The density
of matter after inflation becomes negligibly small, so it must be
created again
in the process of reheating of the universe. This process is strongly
nonadiabatic.

As we already mentioned in the Introduction, in inflationary
cosmology the
bounds of Ref. \cite{lenny3} refer to the post-inflationary particle
horizon,
which means that the integration in Eq. (\ref{parthor}) should begin
not at $t =
0$ or at $t = 1$ but after reheating of the universe. One can easily
verify
that
the bounds
obtained in \cite{lenny3} are valid in this case as well.

\subsection{Closed universe}

The metric of a closed FRW universe is
\begin{equation}
\label{closed} ds^2 = - dt^2 + a^2(t) (d\chi^2 + \sin^2\chi d\Omega) \ ,
\end{equation}
where the spatial part represents a $3$-sphere, with $\chi$
being the azimuthal angle and $d\Omega$ the line element on the polar
$2$-spheres. The lightcones are still bounded by the particle
horizon. However, due to the curvature of the $3$-sphere, the light
rays must now travel along the azimuthal direction in order to
maximize the sphere of causal contact. The comoving horizon is the
extent of the azimuthal angle traveled by light between times $0$ and
$t$:
\begin{equation}
\label{areasph2}\chi_H = {L_H\over a} = \int^t_0 {dt'\over a(t')} \ .
\end{equation}
The boundary area of the
causal sphere is then given by
\begin{equation}
\label{areasph} A \sim 4\pi a^2(t) \sin^2 \chi_H \ .
\end{equation}
The volume inside of this sphere is
\begin{equation}
\label{volsph} V = \int^{\chi_H}_0 d\chi \sin^2 \chi \, d\Omega = \pi
(2\chi_H - \sin 2\chi_H ) \ .
\end{equation}
Assuming a constant comoving
entropy density $\sigma$, we find
\begin{equation} \label{ratio}
\frac{S}{A} = \sigma \frac{2\chi_H - \sin 2\chi_H }{4 a^2(t)
\sin^2 \chi_H } \ .
\end{equation}
Here we have explicitly retained the
contribution from the comoving entropy density $\sigma$, which was
ignored in \cite{lenny3}.

Consider for simplicity a cold dark matter dominated universe, with $p \ll
\rho$. In this case $a = a_{\rm max}
\sin^2(\chi_H/2)$. The moment $\chi_H = \pi$  corresponds to
the maximal
expansion, $a = a_{\rm max}$. But at that time the
light cone
emitted from the ``North pole'' of the universe converges at the ``South
pole,'' the area of the horizon (\ref{areasph2}) vanishes, and the
holographic
bound on the ratio $S/A$ becomes violated \cite{lenny3}. Note that
in all other
respects the point $\chi_H = \pi$ is regular, so one cannot argue,
for example,
that the violation of the holographic bound is a result of violent quantum
fluctuations of the light cone.

\subsection{Open, closed and flat universes with $\Lambda < 0$}

Let us return to the discussion of
the flat universe case and  look at Eq. (\ref{comhorizon2})
again. The size of the comoving horizon $r_H$ can only
grow. Despite this growth, the  holographic bound  is satisfied for
$\rho > 0$, $p > -\rho$,
because the value of $a^2$ grows faster than $r_H$ in this
regime. But this bound can be violated if $a^2$ grows more slowly
than $r_H$, and it will definitely be
violated in all cases where a flat space can collapse.

Usually, cosmologists believe that closed universes collapse,
whereas open or
flat universes expand forever. But the situation is not quite
so simple.
If there is
a sufficiently large positive cosmological constant,
then even a
closed universe will
never collapse.
On the other hand, if  the cosmological
constant is negative, then, even
if it is extremely small, eventually it
becomes dominant, and the universe collapses, independently of
whether it is
closed, open or flat. In all of these cases the holographic
principle,
as formulated in \cite{lenny3}, will be violated.

For simplicity, we will consider a flat universe ($k = 0$) with a negative
vacuum energy density  $-\lambda < 0$, so  that $\rho = p/\gamma
-\lambda$.
We will assume that $\lambda \ll 1$ in  Planckian units. For
example, in our
universe $\lambda$ cannot be greater than $10^{-122}$.  In an expanding
universe $\rho =  \frac{\rho_0}{a^{3(\gamma+1)}} - \lambda$, and the
Friedmann
equation
\begin{equation}
3H^2 = \frac{\rho_0}{a^{3(\gamma+1)}} - \lambda
\label{friedcc}\end{equation}  can be rewritten
as
\begin{equation} \dot a = \pm {1\over \sqrt 3}
\sqrt{\frac{\rho_0}{a^{3\gamma+1}} -
\lambda a^2} \ .
\label{adotcc}
\end{equation}
Because of the presence of
the cosmological term, in general we cannot write the integrals in a
simple form. However, the exact form of the solutions is not necessary
for our purpose here.

First of all, we see that $\dot a$ vanishes at $\lambda a^{3(\gamma
+ 1) }=
\rho_0$, after which $\dot a$ becomes negative and the universe
collapses. This
happens within a finite time after the beginning of
the expansion. From the
definition
of the particle horizon and (\ref{adotcc}), one can find the
value of $L_H$ at the turning point:
\begin{equation}
L_H(turning) =
\frac{B(\frac{\gamma}{2(\gamma+1)},\frac12)}{3(\gamma+1)\sqrt{\lambda}}
\ ,
\end{equation}
where $B(p,q)$ is the Euler beta function. Putting these
formulas together, we see that at the turning point
\begin{equation}
\frac{S}{A} \sim \sigma \lambda^{\frac{1-\gamma}{2(1+\gamma)}}
\end{equation}
up to
factors of order unity. For $1\ge \gamma > -1$,
the power of
$\lambda$ is
positive and so the ratio $S/A$
is very small at the turning point. Now, we can consider what happens
near the final stages of collapse, where the energy density reaches
the Planckian scales. By symmetry, $L_H \sim 2 \frac{a_0}{a(turning)}
L_H(turning)
\sim \lambda^{-(3\gamma+1)/[6(\gamma+1)]}$ at this time, whereas
$\sigma/a^3 \sim 1$. Hence,  Eq.
(\ref{comhorizon2a}) yields
$S/A\sim \lambda^{-(3\gamma+1)/[6(\gamma+1)]}
\gg 1$ when $\gamma>-1/3$.
Therefore, we see that the ratio $S/A$
reaches unity
at some time after the turning point,
and that the holographic bound becomes violated  thereafter,
but still well in the
classical phase, when the universe is still very large. Indeed, we
can estimate the density of matter at that time to be $\rho \sim
\lambda^{\frac{\gamma+1}{2}}\ll 1$.

A universe
where the only energy density is in form of
a negative cosmological constant is called the anti de
Sitter
space (AdS).
In string theory, AdS spaces typically emerge after
compactifying string or M
theory on an internal, compact, Einstein
space of positive constant curvature.
Many interesting
applications
of the holographic principle have been elaborated for the
pure AdS space. It is therefore
quite interesting that in the cosmological context an AdS background
containing
matter describes  a
collapsing Friedmann universe with a
negative vacuum energy, in which the
cosmological holographic principle is violated.

\subsection{AdS spaces with matter and an alternative
formulation of cosmic
holography}

In order to cure the problems of the original formulation of the
cosmological
holographic principle, Bak and Rey proposed a different formulation
\cite{brtwo}. They suggested to consider the so-called apparent
horizon instead
of the particle horizon and claimed that in this case the
holographic bound
holds even in a closed universe. We will not present here a detailed
discussion
of their proposal. Instead we will
consider here their  holographic
bound in
the three-dimensional
spatially flat universe (d = 3), see Eq. (16) of
\cite{brtwo}:
\begin{equation}
{ 4 \sigma \over 3   a^{2}(t)\dot{a}(t) } \quad \le \quad 1 .
\label{flatcondition}
\end{equation}
 This condition is violated when the universe approaches the turning
point at
$\lambda a^{3(\gamma + 1) }= \rho_0$, when one has $\dot a = 0$.
This violation
occurs even much earlier than in the original formulation of the
cosmological
holographic principle of Ref. \cite{lenny3}.

One can propose two possible interpretations of these results. First
of all,
one may argue that closed universes are impossible, and
that the universes with a
negative cosmological constant are also impossible. We do not see
how one could
justify such a statement. After all, the main reason why the holographic
constraint was violated in both cases studied above was related to the
possibility of gravitational collapse. It would be very
odd to
expect
that the holographic principle which was motivated by the study of
black holes
should imply that gravitational collapse cannot occur.

Another possibility is that the formulations of the cosmic holography
proposed in \cite{lenny3,brtwo} should be somewhat modified in the
cases when
the universe may experience collapse.    It would also be
interesting to
understand the
reasons why the holographic inequalities
were correct in the flat universe
case. We will discuss this issue in the next section.

\section{Black holes as big as a universe}

The simplest way to understand the holographic bound on the entropy of the
observable part of the universe is related to the theory of black
holes. In
what follows we will develop  further an
argument given by Easther
and Lowe
\cite{el}, and by Veneziano \cite{gv}.

The simplest cosmological models are based on the assumption that
our universe
is homogeneous. But how do we know that it is indeed homogeneous if
the only
part of the universe that we can see\footnote{If one takes into account
inflation, then particle horizon is exponentially large. Still  we
can  see
(by means of electromagnetic radiation) only a small part of the
universe of
size $\sim H^{-1} \sim t$. It is important that this  scale,  rather
than the
particle horizon, determines the largest size of a black hole which can be
formed in an expanding universe.}  has size $H^{-1}$? We cannot
exclude the
possibility that if we wait for another 10 billion years, we will
see that we
live near
the center of an expanding but isolated gravitational system of size
$O(H^{-1})$ in an asymptotically flat space. Then we can apply the
Bekenstein
bound to the entropy of this system,
$S \lesssim ER$,
where $E \sim \rho
R^3$ is the
total energy and $R \sim H^{-1} $ is the size of this system, with
$H^2 \sim
\rho$, in Planck units. This gives $S \lesssim H^{-2}$, which
coincides with
the holographic bound.

Of course, the idea that our part of the universe is a small
isolated island of
size $H^{-1}$ is weird, but we do not really advocate this view here.
Rather, we
simply say that since we cannot tell  whether the universe is
homogeneous, or it
is an island of a size somewhat greater than $ H^{-1} $, the  bound $S
\lesssim
H^{-2}$ must hold for a usual homogeneous universe as well.

One can look at this constraint from a different perspective. It is
well known
that if our universe is locally overdense on a scale of horizon with
${\delta\rho\over \rho} = O(1)$, the overdense part will collapse
and form a
black hole of a size $H^{-1}$ \cite{carrhawk}.  Then the entropy of
this part
of the universe will satisfy the black hole bound $S \lesssim
H^{-2}$. Again,
there is no indication  that   ${\delta\rho\over \rho} = O(1)$ on
the horizon
scale, but since we cannot exclude this possibility on a scale
somewhat greater
than the present value of $H^{-1}$, the bound should apply to the
homogeneous
universe as well.

Instead of debating the homogeneity of our universe, one can imagine
adding a
sufficient amount of cold dark matter to a part of our universe of
size $R$.
This would not change its entropy, but it would lead to black hole
formation.
Then one can find an upper bound on the entropy of a black hole of
size $R$: $S
\lesssim R^2$.
If one takes
$R \sim H^{-1}$, one again finds that $S \lesssim H^{-2}$.

The bound $S
\lesssim R^2$  implies that the density of entropy satisfies the
constraint $s
= {S/ R^3} < 1/R$. Thus one could expect that it is possible to get a more
stringent constraint on the density of entropy by considering black
holes of
size greater than $H^{-1}$.
 However, according to Carr and Hawking \cite{carrhawk},   black
holes formed
in a flat universe cannot have size   greater than $O(H^{-1})$. This
constraint
has a dynamical origin, and is not related to the size of the
particle horizon.
Usually the difference between $H^{-1}$ and the particle horizon is
not very
large, but during inflation  this difference is very significant: $H^{-1}$
remains nearly constant, whereas the particle horizon grows exponentially.

If  an inflationary domain is homogeneous
on a scale $O(H^{-1})$, then it is going to expand
exponentially,
independently of any inhomogeneities on a larger scale.  Such a
domain is not
going to
collapse  and form a black hole until inflation ends and  we wait
long enough
to see the
boundaries
of the  domain. But this will not happen for an exponentially long
time.
Nevertheless
the holographic constraints on the entropy can be derived for the
processes
after inflation, just as in the case considered above.
These
constraints
will be related to the size of the largest black hole which can be
formed during
the expansion of the post-inflationary universe, $R \sim H^{-1}$,
rather than
to the
exponentially large size of the particle horizon in an inflationary
universe.
As a result, the holographic bounds do not lead to the constraints on the
duration of inflation, inflationary density perturbations, and other
parameters
of inflationary theory discussed in \cite{infl}.

If the universe is non-inflationary and  closed,  or if it has a
negative
cosmological constant,
then, prior to the point of maximal expansion, the holographic
constraints on
the entropy within the
regions of size $H^{-1} \sim t$   coincide with the constraints for the
flat universe
case. Once the universe begins to collapse, the constraints cannot
be further
improved because the typical time of formation of a black hole of
size $O(t)$
at that stage
will be of the same order of magnitude
as the lifetime of the universe. But this
fact does
not imply the impossibility of collapsing universes.

Note that in our consideration we did not make any assumptions about the
adiabatic evolution of the universe. Thus, the cosmological holographic
constraints on entropy are as general as their black hole counterparts. In
fact, we believe that these two constraints
have the same origin.

\section{Holography vs. Inflation}

As we already mentioned,  all holographic constraints  discussed
in this
paper apply only to the post-inflationary universe. Inflationary
cosmology in its spirit is somewhat  opposite to holography. The
possibility of
solving the horizon,
homogeneity,
isotropy, and flatness problems is related to the superluminal
stretching of
the universe, which
erases all memory about the boundary conditions. The speed of
rolling of the
inflaton scalar field approaches an asymptotic value which does not
depend on
its initial speed. The gradients of the fields and the density of
particles
which existed prior to inflation (if there were any) become  exponentially
small. All particles (and all entropy) which exist now in the
universe have
been created after
inflation in the process of reheating. This process occurs locally, so the
properties of particles as well as their entropy do not depend on the
initial conditions in the universe.

In order to investigate this issue in a more detailed way, let us
consider the
simplest version of inflationary cosmology where the universe during
inflation
expands only $10^{30}$ times (the minimal amount which is necessary for
inflation to work). We will also assume for simplicity that
inflation occurs at
the GUT scale, so that $H \sim 10^{-6}$ and the temperature after
reheating is
$T \sim 10^{-3}$ in the Planck units.

In such a case the size of the particle horizon after inflation will
be $L_H
\sim H^{-1} \times 10^{30} \sim 10^{36}$, the area $A \sim L_H^2 \sim
10^{72}$, and
the entropy $S \sim T^3 L_H^3 \sim 10^{99}$, which clearly violates
the bound
$S < A$.
This means that the information stored at the surface of an
inflationary domain cannot describe dynamics
in its interior.

In practice, it is extremely difficult to invent inflationary
theories where
the universe grows only by a factor of $10^{30}$ because typically in such
models ${\delta\rho\over \rho} = O(1)$ at the scale of the horizon. In the
simplest versions of chaotic inflation the universe grows more than
$10^{1000000}$ times during inflation.  The situation becomes especially
dramatic in those versions of inflationary cosmology which lead to
the process
of eternal self-reproduction of inflationary domains. In such models the
universe is not an expanding
ball of a huge size, but a growing fractal consisting of many
exponentially
large balls. In the process of eternal self-reproduction of the
universe all
memory about the boundary conditions and initial conditions becomes
completely
erased \cite{book}.

Of course, one can  use the version of the holographic principle
describing the post-inflationary evolution of the universe, as
discussed in the
previous sections.  However, in realistic inflationary models the energy
density at the end of inflation falls more than 15
orders of magnitude below the Planck density, and the most
interesting part of dynamics of the universe where quantum gravity
could play
a significant role is already over.

There is another interesting aspect of  relations between inflation and
holography.
The holographic bound on the present
entropy of the universe is $S \lesssim H^{-2}$. One has  $H^{-1}
\sim 10^{60}$
 in the Planck units. This gives the constraint
\begin{equation}\label{bound}
S \lesssim H^{-2} \sim 10^{120} \ .
\end{equation}
 Meanwhile, the entropy of matter in the observable part of the
universe is
smaller than $10^{90}$.
If one  thinks about cosmology in terms of the
information which can be stored on the  horizon (or, to be more
accurate, on a
surface of a sphere of size $H^{-1}$), one can be encouraged by the
fact that
the holographic bound is satisfied with a wide safety margin,  $S/A
\lesssim
10^{-30}$. On
the other hand, if, as we have argued, the
information stored on
the sphere of size $H^{-1}$ is not related to the initial conditions
at the
beginning of
inflation, then its importance is somewhat limited. In such a case
the only
information about the universe that we gained is the bound $S \lesssim
10^{120}$, which is $30$ orders of magnitude less precise than the
observational constraint on the entropy. But what is the origin of
these $30$
orders of magnitude?

Let us look back in time and assume that there was no inflation and the
evolution of the universe was adiabatic. Our part of the universe today
has size $\sim 10^{28}$ cm.
At the Planck time its size $l$ would be $10^{28}$ cm
multiplied by $ {T_p\over T_0}$, where $T_0$ is the present value of the
temperature of the universe, and $T_p \sim 1$ is the Planck
temperature. (Note
that the scale of the universe is inversely proportional to $T$ during
adiabatic expansion.)  One therefore
finds  $l \sim 10^{-3}$ cm, which is
$10^{30}$ times
greater than the Planck length. That is exactly the reason why we need the
universe to inflate by the factor of $10^{30}$. (The true number
depends on the
value of reheating temperature after inflation.)

If the universe did not inflate at all, it would be very
holographic. A typical
homogeneous part of the universe soon after the big bang would have Planck
size, it would contain just one or two particles, and the constraint
$S < A$
would be saturated. But we would  be unable to live there.

Let us assume, for the sake of the
argument, that inflation starts and
ends at the
Planck density, and it has Planckian temperature after reheating.  If the
universe during this period inflated by more than $10^{30}$ times,
then our
part of the universe after inflation would have the size $10^{-3}$
cm, i.e.
$10^{30}$ in Planck units, just as we estimated above. Its entropy
would be
$10^{90}$. Then the universe   expands by  $ {T_p\over T_0} \sim 10^{30}$
times, and the area of our domain becomes $10^{120}$. This makes it
clear that
the factor of $10^{30}$ which characterizes the discrepancy between the
holographically natural value of entropy $10^{120}$ and the observed value
$10^{90}$ is the same factor which appears in the formulations of
the entropy
problem and flatness problem \cite{gv}.

Thus, in the final analysis,  the reason why one has  $S \lesssim
10^{-30} A$ in
our  universe is related to inflation.
Without inflation one
would have $S
\sim A$, and a typical locally
homogeneous patch of the universe would collapse within  the Planck
time.
The safety
margin  of 30 orders of magnitude
created by inflation makes the universe very large and long-living, but
simultaneously
prevents the holographic constraint on entropy from
being very informative.

A nontrivial relation between the holographic constraint and
inflation does not
mean that one can identify the entropy problem (existence of a huge
entropy $S \sim 10^{90}$ in our part of the universe) and the holography
problem (discrepancy between the holography bound $10^{120}$ and the
true value
of entropy $10^{90}$). For example, in one of the recent versions of
the pre-big
bang scenario the stage of the pre-big bang inflation begins from a
state which
can be identified with a black hole with a large  area of the black hole
horizon \cite{damour}.
In this case, the initial entropy of the gravitational
configuration by definition satisfies the Bekenstein-Hawking bound, which
coincides
with the holographic bound.
If one assumes that the entropy of
matter inside
the black hole {\it saturates} the Bekenstein-Hawking bound (this is
just an
assumption which does not follow from the black hole theory),
then the holography problem will be resolved \cite{gv}.
However, one should still
determine the
origin of the enormously large black
hole entropy in this scenario, which
constitutes the entropy problem  \cite{klb}.

\section{Conclusions}
The idea that all information about physical processes in the world can be
stored
on its surface is very powerful. It has many interesting implications in
investigation of the nonperturbative properties of M-theory.
However, it is
rather difficult to merge this idea with cosmology. The universe may
not have
any boundary at all, or it may expand so fast that  boundary effects
become
irrelevant for the description of the local dynamics.
In this paper we have shown that some of the formulations of the
holographic
principle should be modified not only in application to a closed
universe, but
also for open, closed and flat universes with a negative
cosmological constant.
We believe that the cosmological holographic constraints on entropy,
in those
cases where they are valid, can be understood using the Bekenstein-Hawking
bound on the entropy of black holes. These constraints are rather
nontrivial,
but if applied
to our part of the universe they are much weaker
than the
observational constraints, as well as the constraints which follow
from the
theory of creation of matter after inflation. We believe that these
constraints
do not permit one
to rule out the universes which may experience
gravitational
collapse, and they do not impose any additional constraints on
inflationary
cosmology.

The constraints on entropy represent only one aspect of the holographic
principle.
A stronger form which has been advocated
requires the existence of a theory
living on the boundary surface which would
describe physical processes in the enclosed volume.
Validity of this conjecture in the cosmological context has
not been
demonstrated, and in fact one may argue that there exists a general
obstacle on
the way towards the realization of this idea. In the theory of black
holes, the role
of the holographic surface is played by the black hole horizon. Its
area, and
correspondingly the number of degrees of freedom living on the
horizon, remains constant if one neglects quantum gravity effects.
Thus it is not
unreasonable
to assume that there exists a unitary quantum theory associated with
the black hole horizon.
However,
in an expanding universe the number of degrees of freedom
associated with the cosmological horizon, or with apparent horizon,  
or with a
horizon of a
would-be black
hole which provides holographic constraints on entropy,  rapidly
changes in
time. For example, in a closed universe the initial area
of the horizon is
vanishingly small, then it grows until it reaches the
maximum, and subsequently it disappears. Thus the
number of degrees of freedom associated
with such a surface strongly depends on time   even when the  
evolution of the
universe is adiabatic and the total number of degrees of freedom in  
the bulk is
conserved  \cite{RB}.

Therefore one may wonder whether the holographic theory existing on such a
surface will violate unitarity. In addition, the disappearance of  
degrees of
freedom after the moment of the maximal expansion implies that  the  
entropy
measured at the holographic surface will increase during the universe
expansion, but then it will decrease during its contraction, and  
eventually it
will vanish. This means that the second law of
thermodynamics may be violated in the holographic theory.

 The situation with causality in such a theory is not clear as well.
Indeed,
information about the new degrees of freedom which are going to appear or
disappear on the
holographic surface is stored not on this surface but in the bulk. This
information does not
propagate
along the surface, rather it crosses the surface when new particles
enter the
apparent horizon.  But this suggests that the creation of the new
degrees of
freedom in the holographic theory will not look like an effect
caused by the
earlier existing conditions at the surface.

It remains to be seen whether one can overcome all of these
problems  and
make the
holographic principle a useful part of the modern cosmological
theory including
inflationary theory.
We should note, however, that  quantum cosmology is extremely
complicated and
counterintuitive in many respects. It is still a challenging task to unify
M-theory and inflationary cosmology. Any progress in this direction
would be
very important. One may expect that the ideas borne out by
the investigation of
quantum dynamics of black holes and enriched by the study of
supergravity and
string theory will play the key role in the development of a
nonperturbative
approach to quantum cosmology.

\

 We wish to thank R. Bousso, W. Fischler and L. Susskind for valuable
discussions. This work has been supported in part  by NSF grant
PHY-9870115.

\end{document}